\documentclass[prc,preprint,tightenlines,showpacs]{revtex4}

\usepackage{graphicx}  
\usepackage{bm}  
\usepackage{amsmath}

\newcommand{\beq}{\begin{equation}}
\newcommand{\eeq}{\end{equation}}
\newcommand{\bea}{\begin{eqnarray}}
\newcommand{\eea}{\end{eqnarray}}

\begin{document}
\preprint{\tiny{FZJ-IKP-TH-2005-15}}
\preprint{\tiny{HISKP-TH-04/19}}
\preprint{\tiny{INT-PUB 04-24}}
\title{On the correlation between the binding energies of the 
 triton and the $\alpha$-particle}
\author{L. Platter}\email{l.platter@fz-juelich.de}
\affiliation{Forschungszentrum J{\"u}lich, Institut f{\"ur} Kernphysik,
    D-52425 J{\"u}lich, Germany}
\affiliation{Helmholtz-Institut f\"ur Strahlen- und Kernphysik (Theorie),
Universit\"at Bonn, Nu\ss allee 14-16, D-53115 Bonn, Germany}
\author{H.-W. Hammer}\email{hammer@phys.washington.edu}
\affiliation{Institute for Nuclear Theory, University of Washington, Seattle, 
WA 98195, USA}
\author{Ulf-G. Mei{\ss}ner}\email{meissner@itkp.uni-bonn.de}
\affiliation{Helmholtz-Institut f\"ur Strahlen- und Kernphysik (Theorie),
Universit\"at Bonn, Nu\ss allee 14-16, D-53115 Bonn, Germany}
\affiliation{Forschungszentrum J{\"u}lich, Institut f{\"ur} Kernphysik,
    D-52425 J{\"u}lich, Germany}
\date{\today}
\begin{abstract}
We consider the correlation between the binding energies of the 
triton and the $\alpha$-particle which is empirically observed
in calculations employing different phenomenological nucleon-nucleon
interactions. Using an effective quantum mechanics approach
for short-range interactions with large scattering length $|a|\gg \ell$,
where $\ell$ is the natural low-energy length scale,
we construct the effective interaction potential at leading order in 
$\ell/|a|$.
In order to renormalize the four-nucleon system, it is sufficient to include
a $SU(4)$-symmetric one-parameter three-nucleon interaction in addition to
the S-wave nucleon-nucleon interactions. The absence of a four-nucleon
force at this order explains the empirically observed correlation between the 
binding energies of the triton and the $\alpha$-particle. 
We calculate this correlation and obtain a prediction for the $\alpha$-particle
binding energy. Corrections to our results are suppressed by $\ell/|a|$.
\end{abstract}
\pacs{21.45.+v, 21.10.Dr, 03.65.Ge}
\maketitle
The scattering of two particles with short-range interactions 
at sufficiently low energy is determined by their S-wave scattering length
$a$. Generically, the scattering length $a$ is comparable to the 
typical low-energy length scale 
of the system $\ell$ which,  for short-range interactions,
is of the order of the range of the potential. 
In special cases, however, the scattering length can be much larger,
$|a|\gg \ell$, due to a fine-tuning of the parameters in the underlying 
interaction potential. Such a fine-tuning can be accidental 
as in the case of two nucleons at low energy, or it can be controlled 
experimentally by varying an external parameter such as a 
magnetic field in the case of Feshbach resonances for atomic systems.

Few-body systems with large scattering length are particularly
interesting because such systems show many universal properties. 
The simplest example for positive $a$
is the existence of a shallow bound state (dimer) with binding energy
\beq
B_d=\frac{\hbar^2}{ma^2}+ {\cal O}(\ell/a)\,, \qquad a>0\,,
\label{B2-eq}
\eeq
where $m$ is the mass of the particles.
The separation of scales between $|a|$ and $\ell$ also manifests itself
in few-body systems with more than two particles. The most prominent
example is the Efimov effect: the occurence of a geometric spectrum of 
three-body bound states in the limit $a\to \pm\infty$ 
\cite{Efimov71}. In this limit, the ratio of the binding energies of
successive Efimov states approaches the universal number 
$515.03..\,$. In order to fully specify the bound state spectrum,
however, the absolute energy of one Efimov state has to be specified
and this introduces one three-body parameter, $L_3$, which 
will be discussed in more detail below. As a consequence,
a three-body system with a large but finite scattering length can be 
characterized by specifying the values of $a$ and $L_3$.

There has been a considerable interest in physical systems with 
large scattering length recently. 
This interest was stimulated by the experimental realization of 
Feshbach resonances with trapped atoms and by the effective 
field theory program for nuclear physics. (See, e.g., 
Refs.~\cite{Braaten:2004rn,Bedaque:2002mn} and references therein.)
The best known example of a nuclear system with a large scattering
length is the two-nucleon ($NN$) system. There are two independent S-wave
scattering lengths that govern the low-energy scattering 
of nucleons. The scattering lengths 
$a_s= -23.5$ fm and $a_t= 5.42$ fm describe $NN$ scattering in the 
spin-singlet ($^1S_0$) and spin-triplet ($^3S_1$) channels, respectively.
Both scattering lengths are significantly larger than the natural
low-energy length scale $\ell\sim\hbar/(m_\pi c)\approx 1.4$ fm, while
the effective ranges are of the same order as $\ell$.
As a consequence, an expansion in $\ell/|a|$ is useful. 

A peculiar universal feature of the three-nucleon system is the Phillips 
line \cite{Phillips68}. If the predictions of different nucleon-nucleon
potentials for the triton binding energy $B_t$ and the spin-doublet
neutron-deuteron scattering length  $a_{nd}^{(1/2)}$ are plotted
against each other, they fall close to a line.
This correlation between $B_t$ and $a_{nd}^{(1/2)}$ is called the 
Phillips line and can not be understood in conventional potential
models. However, it immediately follows from universality if the 
large $NN$ scattering lengths are exploited within an expansion in  
$\ell/|a|$ \cite{EfiTk85,Bedaque:1999ve}. 
If corrections of order $\ell/|a|$ are neglected, all low-energy
3-nucleon observables depend only on the spin-singlet and spin-triplet
scattering lengths $a_s$ and $a_t$ and the three-body parameter $L_3$.
Since the $NN$ potentials reproduce the 
scattering phase shifts, they all have the same scattering
lengths. The short distance part of the potentials which is encoded
in the three-body parameter $L_3$, however, is
not constrained by the phase shifts and in general is different for 
each potential. The different potential model calculations must therefore
fall close to a line which is parametrized by the 
parameter $L_3$.
A similar universal feature of the four-nucleon system is
the Tjon line: an approximately linear correlation between the 
triton binding energy $B_t$ and the binding energy of the $\alpha$-particle
 $B_\alpha$. This correlation was discovered
by Tjon \cite{Tjo75} using simple separable interactions,
but also holds for modern phenomenological potentials \cite{Nogga:2000uu}. 
The origin of this correlation has not been explained. However, it
would immediately follow if the four-nucleon system could be renormalized
to leading order in $l/|a|$ without specifying a four-body parameter.
The purpose of this paper is to study the four-nucleon system 
in an effective quantum mechanics approach (see, e.g.,
Ref.~\cite{Lepage:1997cs}) 
to leading order in $\ell/|a|$ in order to understand the origin of the 
Tjon line. We will show that no new four-body parameter is required
to renormalize the four-nucleon system and calculate the binding
energy of the $\alpha$-particle to leading order in $\ell/|a|$. 

Effective theories are a convenient tool to calculate observables of low-energy
systems to high precision in a systematic fashion.  They are ideally
suited to exploit a separation of scales such as the one between $|a|$
and $\ell$. 
At sufficiently low energy, every interaction can be considered pointlike. One
can then use a very general effective theory to describe the universal
low-energy properties of the system in which observables are computed with
a controlled expansion in $\ell/|a|$.
Few-body systems with large scattering length are of special interest:
It is known that in the three-body sector with short-range interactions a 
three-body force at leading order is needed to renormalize the theory 
\cite{Bedaque:1998kg}.  This implies that at leading order
in $l/|a|$, the properties of the three-body system with large scattering
length are not determined by two-body data alone and one piece of
three-body information (such as the three-body binding energy) is required
as well.
Recently we considered the four-boson system with short-range interactions. We
showed that at leading order no four-body force is needed for 
consistent renormalization and applied the theory to atomic $^4$He
clusters \cite{Platter:2004qn}. 

In this letter, we extend this work to the four-nucleon system.
There has been a large interest recently in applying effective field theory
methods to the few-nucleon problem.
(See, e.g., Refs.~\cite{Bedaque:2002mn,Meissner:2004yy,Navratil:2003ib} 
and references therein.) 
The four-nucleon problem is more complicated than the four-boson problem due
to spin and isospin. Both the $^1S_0$ and $^3S_1$ $NN$ scattering lengths
$a_s$ and $a_t$ contribute at leading order in $\ell/|a|$. 
However, as in the case of bosons, only one three-body parameter $L_3$
enters for the $\alpha$ particle.
The parameter $L_3$ determines the $SU(4)$-symmetric 
three-nucleon force that is required to renormalize the three-nucleon 
problem \cite{Bedaque:1999ve}.
In the following, we will show that no four-body parameter is required
to renormalize the nuclear four-nucleon system and that
the binding energy of the $\alpha$-particle can be described to 10\%
accuracy at leading order in $\ell/|a|$. 
Furthermore, we will calculate the Tjon line at leading order 
in $\ell/|a|$.

The effective low-energy interaction potential generated by 
short-range contact interactions can be written
down in a momentum expansion. In the S-wave sector of the two-nucleon
system, it takes the general form
\beq
\langle {\bf k'} | V | {\bf k} \rangle =
{\cal P}_s\,\lambda_2^s\, g({\bf k'})g({\bf k}) +{\cal P}_t\,\lambda_2^t\,
g({\bf k'})g({\bf k}) +\ldots\,,
\label{effpot}
\eeq
where the dots indicate momentum dependent terms that are higher
order in $\ell/|a|$ and ${\cal P}_s$ and ${\cal P}_t$ project onto the 
$^1S_0$ and $^3S_1$ partial waves, respectively.
Because of Galilean invariance, the interaction can only depend on the 
relative momenta of the incoming and outgoing particles 
${\bf k}$ and ${\bf k'}$. The regulator function 
$
\langle{\bf u}|g\rangle= g({\bf u})=\exp(-{u^2}/{\Lambda^2})\,
\label{eq-regu}
$
cuts off the contribution of momentum states with $u \gg \Lambda$.
The cutoff parameter $\Lambda$ is arbitrary and in the end all
observables should be independent of $\Lambda$.
The interactions in Eq.~(\ref{effpot})
are separable and thus, the two-body problem
for each partial wave can be solved analytically. The two-body
t-matrix can be written as 
$t_{t,s}(E)=|g\rangle\tau_{t,s}(E)\langle g|$,
where $E$ denotes the energy. The two-body propagator $\tau_{t,s}(E)$ is
given by:
\begin{equation}
\tau_{t,s}(E)=\Bigl[1/\lambda_2^{t,s}-4\pi\int\hbox{d}q\,
  q^2\frac{g(q)^2}{E-q^2}\Bigr]^{-1}~.
\label{eq-tau}
\end{equation}
Here and in the following we set $\hbar=m=1$. 
The coupling constants $\lambda_2^s$ and $\lambda_2^t$ can be fixed by 
demanding that the triplet and singlet scattering length $a_t$ and
and $a_s$ 
are reproduced correctly by the corresponding t-matrices.

The properties of the triton are determined by the Faddeev
equations. 
To leading order in $\ell/|a|$, all internal orbital angular 
momenta can be set to zero. Thus, the triton spin $1/2$ is built 
up from the spins of the nucleon only.
The triton wave function can be decomposed into Faddeev components.
Since we are mainly interested in the binding energies and not in the
wave functions, we can eliminate all but one of the components and obtain
an equation for remaining component $\psi$ :
$
\psi=G_0 t P \psi+G_0 t G_0 t_3 (1+P)\psi\,,
$
where $G_0$ is the free three-particle propagator and
$P=P_{13}P_{23}+P_{12}P_{23}$ is a permutation operator that 
generates the omitted Faddeev components; $P_{ij}$ exchanges
particles $i$ and $j$. The auxilliary quantity
$t_3$ is the solution of a Lippmann-Schwinger equation with 
a $SU(4)$-symmetric three-body contact interaction 
\beq
V_3={\cal P}_a \lambda_3 h(u_1,u_2) h(u'_1,u'_2)
\eeq
only. Here ${\cal P}_a$ denotes the projector on the total antisymmetric 
three-body state with total spin $S=1/2$ and total isospin $T=1/2$
as for example given in \cite{Delfino:1984}.
The regulator function $h(u_1,u_2)=\exp(-(u_1^2+{\textstyle \frac{3}{4}}u_2^2)/
\Lambda^2)$ is defined in terms of the familiar Jacobi momenta
of the three-body system. 
The three-body coupling $\lambda_3$ must be determined from a three-body
observable. This interaction is required in order to renormalize the 
three-body system and achieve independence of the cutoff
parameter $\Lambda$.
Its renormalization group behavior is governed
by a limit cycle \cite{Bedaque:1999ve,Platter:2004qn}.
For large values of $\Lambda$ the running of the coupling
constant is described by
\beq
\lambda_3(\Lambda)=\frac{c}{\Lambda^4} \;\frac{\sin(s_0 \ln(\Lambda/L_3)
  -\arctan(1/s_0))}{\sin(s_0 \ln(\Lambda/L_3)+\arctan(1/s_0))}~,
\label{eq:limcyc}
\eeq
where $c\approx 0.016$ is a normalization constant and
$s_0\approx 1.00624$ is a transcendental number that determines the
period of the limit cycle. If the cutoff $\Lambda$ is multiplied by 
a factor $\exp(n\pi/s_0) \approx (22.7)^n$ with $n$ an integer, the 
three-body coupling $\lambda_3$ is unchanged.
The three-body parameter $L_3$ can be determined directly from observable 
quantities like the triton binding energy $B_t$. 

Having expressed all relevant coupling constants in terms of 
physical observables, we are now ready to compute the binding
energies of the four-nucleon system. 

The binding energies of a four-body system of nucleons can 
be determined with the Yakubovsky equations. 
The four-body wavefunction $\Psi$ is first decomposed into Faddeev 
components, followed by a second decomposition into so-called
Yakubovsky components. In the case of nucleons
one ends up with two components $\psi_A$ and $\psi_B$. 
In the following, we will consider the $\alpha$-particle
with total angular momentum $J=0$ and total isospin
$T=0$. The Yakubovsky equations can be written as \cite{Glockle:1993vr}:
\bea
\psi_A&=&G_0 t P \Bigl[(1-P_{34})\psi_A + \psi_B\Bigr]
+\frac{1}{3}(1+G_0 t_{12})G_0 V_3 \Psi\,,
\nonumber\\
\psi_B&=&G_0 t \tilde{P} \Bigl[(1-P_{34})\psi_A + \psi_B\Bigr]\,,
\label{eq-yaku}
\eea
where $G_0$ is the free four-particle propagator and
$\Psi$ is the full four-body wave function that can be
reconstructed from $\psi_A$ and $\psi_B$. 
We solve Eqs.~(\ref{eq-yaku}) numerically in momentum space. 
For details of the solution and structure of the above equations, 
we refer the reader to  Refs.~\cite{Platter:2004qn,PlatterNew}.
The two-body couplings $\lambda_2^t$
and $\lambda_2^s$ are fixed by matching Eq.~(\ref{eq-tau}) to
reproduce the triplet and singlet scattering lengths 
$a_t$ and $a_s$, respectively. 
Instead of $a_t$ one can also use the deuteron binding energy 
$B_d=2.22$ MeV together with Eq.~(\ref{B2-eq}). The difference between 
these two methods is higher order in $\ell/|a|$ and thus gives a
naive estimate of the error made in the order considered.
The three-body parameter $L_3$ which is implicit in the three-body
interaction $V_3$, is fixed from the triton binding energy
$B_t=8.48$ MeV. We find $L_3=21.2$~fm$^{-1}$ ($L_3=19.8$ fm$^{-1}$) 
if $a_t$ ($B_d$) are used 
as input.

We now calculate the $\alpha$-particle binding energy $B_\alpha$
for different values of the cutoff parameter $\Lambda$. We vary
$\Lambda$ in the region where the three-nucleon system has exactly
one bound state. 
For values of $\Lambda$ below  4 fm$^{-1}$ there is a strong cutoff
dependence. For values in the range 8~fm$^{-1}\leq \Lambda
\leq 10$ fm$^{-1}$, however, our results are numerically stable and 
independent of $\Lambda$ to within 5\%. 
This variation is considerably smaller than 
the errors from higher orders in $\ell/|a|$ of about 30\%.
The existence of the plateau
for $\Lambda \geq 8$ fm$^{-1}$ provides sufficient evidence that the 
renormalization of the four-nucleon system at leading order in
$\ell/|a|$ does not require a four-nucleon force. Renormalization
of the the three-nucleon system automatically guarantees
renormalization of the four-nucleon system.

For the $\alpha$-particle binding energy, we find 
$B_\alpha = 29.5$~MeV. If the deuteron binding energy $B_d$
 is used as input instead of the triplet scattering
length $a_t$, we obtain $B_\alpha = 26.9$ MeV. This variation 
is consistent with the expected 30\% accuracy of a leading 
order calculation in $\ell/|a|$. Our results agree
with the (Coulomb corrected) experimental value 
$B_\alpha^{exp}=(29.0 \pm 0.1)$~MeV to within 10\%. We conclude 
that the ground state energy of the $\alpha$-particle can be described
by an effective theory with short-range interactions only.
Of course, a more refined analysis should include Coulomb
corrections. However, the size of these corrections is expected
to be smaller than the NLO contribution.
Furthermore, corrections due to isospin violation will appear naturally
at higher order within the effective theory.

\begin{figure}[tb]
\centerline{\includegraphics*[width=12cm,angle=0]{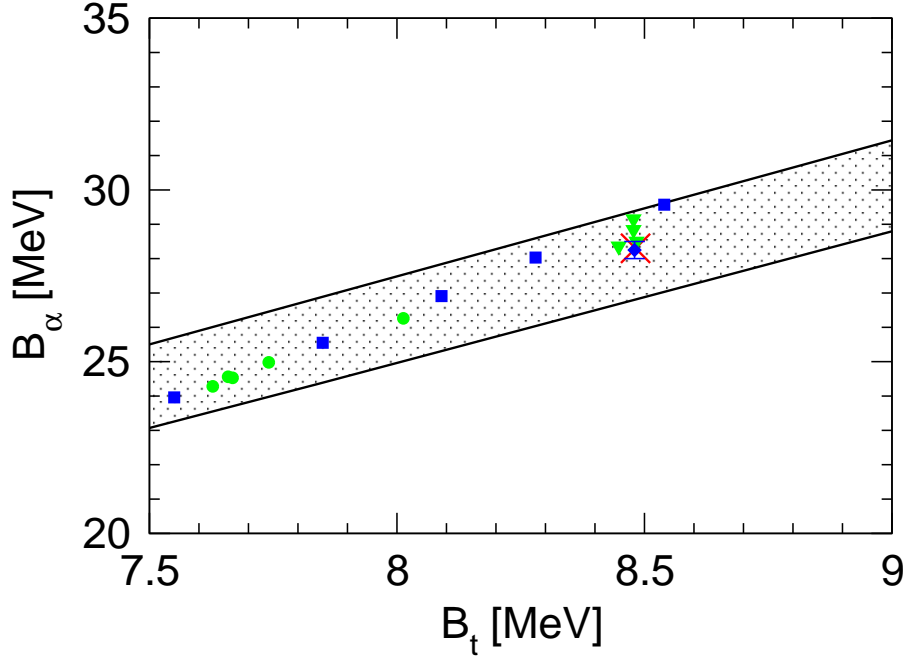}}
\caption{\label{fig:tjon}
The correlation between the binding energies of the 
triton and the $\alpha$-particle (the Tjon line). The lower (upper) line
shows our leading order result using  $a_s$ and $B_d$
($a_s$ and $a_t$) as two-body input. The grey dots and triangles show
various calculations using phenomenological potentials
without or including three-nucleon forces, respectively.
The squares show the results of chiral EFT at NLO for different cutoffs
while the diamond shows the N$^2$LO result. The cross shows the experimental 
point.
}
\end{figure}

The empirically observed correlation between the binding energies of the 
triton and the $\alpha$-particle (the Tjon line) follows 
immediately from the absence of a four-nucleon force at leading
order in $\ell/|a|$. The Tjon line is generated 
by variation of the three-body parameter $L_3$, which 
is different for different $NN$ potentials.
In Fig.~\ref{fig:tjon}, we show our result for the Tjon line with $a_s$
and $B_d$ as input (lower line)
and $a_s$ and $a_t$ as two-body input (upper line). Both lines generate
a band that gives a naive estimate of higher order corrections
in $\ell/|a|$. We also show some calculations using phenomenological 
potentials \cite{Nogga:2000uu} and a chiral EFT potential with explicit
pions \cite{Epelbaum:2002vt,Epelbaum:2000mx}.
The Tjon line is a general feature of few-body systems 
with large scattering length. As a consequence, all calculations with 
interactions that give a
large scattering length should lie close to this line. Different
short-distance physics and/or cutoff dependence should only move 
the results along the Tjon line. This can for example be observed in
the NLO results with the chiral potential indicated by the squares
in Fig.~\ref{fig:tjon} or in
the few-body calculations with the low-momentum $NN$ potential 
$V_{{\rm low} k}$ carried out in Ref.~\cite{Nogga:2004ab}. The 
$V_{{\rm low} k}$ potential
is obtained from realistic $NN$ interactions by intergrating out 
high-momentum modes above a cutoff $\Lambda$
but leaving two-body observables (such as the large
scattering lengths) unchanged. The results of few-body calculations 
with $V_{{\rm low} k}$ alone are not independent of $\Lambda$ but lie all
close to the Tjon line (cf. Fig.~2 in Ref.~\cite{Nogga:2004ab}). (For
a different approach to calculate few-nucleon systems based on a low
momentum potential, see Ref.~\cite{Fujii:2004dd} and references therein.)
A similar conclusion holds for the study with various nonlocal potentials
in \cite{Lazauskas:2004hq}.
Our result for the Tjon line to leading order in $\ell/|a|$ is 
shifted from the empirical values because we neglect finite range effects.
Furthermore, the empirical Tjon line becomes a narrow band because of higher 
order three-nucleon force effects.


In summary, we have computed the $\alpha$-particle binding energy at leading
order in the effective theory with contact interactions.
Our results are in good agreement with the experimental value
and agree within the expected error bounds given by an estimate
of higher order contributions.
At leading order no four-body force is needed for renormalization
and thus, the Tjon line can be understood as a general feature of
low-energy systems with large scattering length. 
As a consequence, all calculations with interactions that give a
large scattering length should lie close to this line. 
Finally, further effort should be devoted to the computation of scattering
observables and the inclusion of effective range corrections.

\begin{acknowledgments}
We thank A.~Nogga for valuable discussions and comments on the manuscript.
L.~Platter thanks the INT in Seattle for its hospitality during
completion of this work. This research was supported in part by the 
German Academic Exchange Service (DAAD), the US Department of
Energy under grant DE-FG02-00ER41132, and the Deutsche Forschungsgemeinschaft
through funds provided to the SFB/TR~16.
\end{acknowledgments}


\begin{thebibliography}{99}

\bibitem{Efimov71}
V.~Efimov,
Sov.\ J.\ Nucl.\ Phys. {\bf 12}, 589 (1971).

\bibitem{Braaten:2004rn}
E.~Braaten and H.-W.~Hammer,
arXiv:cond-mat/0410417.


\bibitem{Bedaque:2002mn}
P.~F.~Bedaque and U.~van Kolck,
Ann.\ Rev.\ Nucl.\ Part.\ Sci.\  {\bf 52}, 339 (2002)
[arXiv:nucl-th/0203055].


\bibitem{Phillips68}
A.C.~Phillips, 
 Nucl.\ Phys.\ A {\bf 107}, 209 (1968).

\bibitem{EfiTk85}
V.\ Efimov and E.G.\ Tkachenko,
Phys.\ Lett.\ {\bf 157B}, 108 (1985);
Sov.\ J.\ Nucl.\ Phys. {\bf 47}, 17 (1988).

\bibitem{Bedaque:1999ve}
P.~F.~Bedaque, H.-W.~Hammer and U.~van Kolck,
Nucl.\ Phys.\ A {\bf 676}, 357 (2000)
[arXiv:nucl-th/9906032].

\bibitem{Tjo75}J.A.~Tjon, 
Phys.\ Lett.\ B {\bf 56}, 217 (1975).

\bibitem{Nogga:2000uu}
A.~Nogga, H.~Kamada and W.~Gl{\"o}ckle,
Phys.\ Rev.\ Lett.\  {\bf 85}, 944 (2000)
[arXiv:nucl-th/0004023].

\bibitem{Lepage:1997cs}
G.~P.~Lepage,
arXiv:nucl-th/9706029.

\bibitem{Bedaque:1998kg}
P.~F.~Bedaque, H.-W.~Hammer and U.~van Kolck,
Phys.\ Rev.\ Lett.\  {\bf 82}, 463 (1999)
[arXiv:nucl-th/9809025];
Nucl.\ Phys.\ A {\bf 646}, 444 (1999)
[arXiv:nucl-th/9811046].


\bibitem{Platter:2004qn}
L.~Platter, H.-W.~Hammer and U.-G.~Mei\ss ner,
Phys.\ Rev.\ A {\bf 70}, 052101 (2004)
[arXiv:cond-mat/0404313].

\bibitem{Meissner:2004yy}
U.-G.~Mei{\ss}ner,
arXiv:nucl-th/0409028.


\bibitem{Navratil:2003ib}
P.~Navratil and E.~Caurier,
Phys.\ Rev.\ C {\bf 69}, 014311 (2004)
[arXiv:nucl-th/0311036].


\bibitem{Delfino:1984}
A.~Delfino and W.~Gl\"ockle,
Phys.\ Rev.\ C {\bf 30}, 376 (1984).

\bibitem{Glockle:1993vr}
W.~Gl\"ockle and H.~Kamada,
Nucl.\ Phys.\ A {\bf 560}, 541 (1993).


\bibitem{PlatterNew}
L.~Platter, H.-W.~Hammer and U.-G.~Mei\ss ner,
in preparation.

\bibitem{Epelbaum:2000mx}
E.~Epelbaum, H.~Kamada, A.~Nogga, H.~Witala, W.~Gl\"ockle and U.-G.~Mei\ss ner,
Phys.\ Rev.\ Lett.\  {\bf 86}, 4787 (2001)
[arXiv:nucl-th/0007057].

\bibitem{Epelbaum:2002vt}
E.~Epelbaum, A.~Nogga, W.~Gl\"ockle, H.~Kamada, U.-G.~Mei\ss ner and H.~Witala,
Phys.\ Rev.\ C {\bf 66}, 064001 (2002)
[arXiv:nucl-th/0208023].

\bibitem{Nogga:2004ab}
A.~Nogga, S.~K.~Bogner and A.~Schwenk,
arXiv:nucl-th/0405016.

\bibitem{Fujii:2004dd}
S.~Fujii, E.~Epelbaum, H.~Kamada, R.~Okamoto, K.~Suzuki and W.~Gl\"ockle,
Phys.\ Rev.\ C {\bf 70}, 024003 (2004)
[arXiv:nucl-th/0404049].

\bibitem{Lazauskas:2004hq}
R.~Lazauskas and J.~Carbonell,
Phys.\ Rev.\ C {\bf 70}, 044002 (2004)
[arXiv:nucl-th/0408048].

\end{thebibliography}
\end{document}